\documentclass{article}
\usepackage{graphicx}
\usepackage{dcolumn}
\usepackage{bm}
\begin{document}
\setcounter{page}{1}
\thispagestyle{empty}
\renewcommand{\thesection}{\Roman{section}}
\begin{center}
{\Large\bf 
Thermodynamic properties of ultracold Bose gas: transition exponents and universality}\\[1cm]
\vspace{0.5cm}

{\normalsize{\it{
Sanchari Goswami, Tapan Kumar Das }}}\\
Department of Physics, University of Calcutta, 92 A.P.C. Road, 
Calcutta-700009, India\\
and \\
{\it Anindya Biswas}\\
Harish-Chandra Research Institute, Chhatnag Road, Jhunsi, Allahabad 211 019, India.\\[1.5 cm]

\end{center}
\vskip 1cm
\begin{center}
{\bf ABSTRACT}\\
\end{center}
We report exact numerical calculation of chemical potential, condensate 
fraction and specific heat of $N$ non-interacting bosons confined 
in an isotropic 
harmonic oscillator trap in one, two and three dimensions, as also for interacting bosons in a 3D trap. 
Quasi phase transitions are observed in all these cases, including one-dimension, as shown by a rapid 
change of all the thermodynamic quantities at the transition point. The change becomes more rapid as $N$ 
increases in 2D and 3D cases. However with increase in $N$, the sudden change in the nature of specific 
heat, gets gradually wiped out in 1D, while it becomes more drastic in 2D and 3D. The sudden change in the 
nature of condensate fraction and chemical potential as $N$ increases becomes more drastic even in 1D. 
Defining transition exponents, which characterize the nature of a thermodynamic quantity at the transition 
point of a quasi phase transition, we evaluate them by 
careful numerical calculation very near the transition temperature. These exponents are found to 
be independent of the size of 
the system and whether the bosons are interacting or not, demonstrating 
their universality property.     \\[1cm] 


\section{Introduction}
\hspace*{1cm}
Bose-Einstein condensation (BEC) is the transition process, in which a macroscopic 
fraction of bosons goes into the lowest energy state, below a certain critical 
temperature ($T_c$)~\cite{Huang}. It was predicted by Einstein in 1925, based on Bose's 
explanation of black body radiation and experimentally observed in harmonically 
trapped bosonic alkali atoms in 1995 (for details see Ref.~\cite{Dalfovo}). 
BEC is generally considered as a {\it phase transition} in the 
thermodynamic limit, {\it i.e.} when the number of bosons ($N$) goes to infinity. 
At a phase transition, thermodynamic quantities usually display a critical behavior. Hence it is 
essential to know how such quantities change at and across $T_c$. Unfortunately, their analytic expressions -- 
known over many decades -- are obtained under several idealized simplifying conditions, which are far from the 
actual laboratory situation. In some cases, the idealized treatment even gives wrong answers, as we will 
see below. Attempts have been made to partly relax the simplifying idealizations and use mathematical 
approximations to obtain analytic or semi-analytic expressions for thermodynamic quantities~\cite{Bagnato,Yukalov,Ketterle}. But a fully 
satisfactory analytic treatment is not possible, making a numerical investigation essential. In this 
communication, we present a numerical study of the critical behavior just across the BEC transition. \\

In standard text books~\cite{Huang,Pathria}, it is shown 
that BEC is possible in an ideal ({\it i.e.} non-interacting) and  homogeneous 
({\it i.e.} not confined) Bose gas below a finite critical temperature 
$T_c^0$ (we use the superscript $0$ to indicate the ideal case) in three and higher dimensions, while
in two-dimension the critical temperature becomes zero and no BEC transition is 
possible in one-dimension. For a three-dimensional ideal and uniform Bose gas 
in the thermodynamic limit ($N \rightarrow \infty$), the condensate fraction 
(defined as $\frac{N_0}{N}$, where $N_0$ is the number of bosons in the lowest 
energy state) and the heat capacity, at a temperature $T< T_c^0$, are given by~\cite{Huang,Pethick}
\begin{eqnarray}
\frac{N_0}{N}=1-{\Big(}\frac{T}{T_c^0}{\Big)}^{\frac{3}{2}},\nonumber \\
C_N(T)=\frac{15}{4}Nk_B\frac{\zeta(\frac{5}{2})}{\zeta(\frac{3}{2})}{\Big(}\frac{T}{T_c^0}{\Big)} ^{\frac{3}{2}},
\label{cf}
\end{eqnarray}
where $k_B$ is the Boltzmann constant and $\zeta(x)=\sum_{n=1}^{\infty} n^{-x}$ is 
the Riemann zeta function. It is assumed that the chemical potential ($\mu$) remains zero for 
$T \leq T^0_c$, and starts to decrease above $T^0_c$. 
Above $T_c^0$, condensate fraction vanishes, 
while the heat capacity takes an involved expression (see Eq.~(12.66) of Ref~\cite{Huang}). 
As functions of temperature, both condensate fraction and $C_N$ are continuous, while their first
derivatives are discontinuous at $T=T_c^0$ (see Fig. 12.9 of Ref.~\cite{Huang}). This means that
in this limit, BEC in an ideal and uniform Bose gas is a continuous 
phase transition process. The situation is quite different if the Bose gas is 
inhomogeneous, {\it i.e.}, confined in a suitable trap~\cite{Pitchard}. Commonly 
used experimental traps are spherical or axially symmetric harmonic oscillator traps, for which 
$\frac{N_0}{N}=1-{\Big(}\frac{T}{T_c^0}{\Big)}^{3}$ and $C_N(T)$ has a {\it finite} discontinuity 
at $T=T_c^0$~\cite{Pethick, Huang, Pitchard} in the thermodynamic limit. Thus, the corresponding 
phase transition is of first order. \\

These analytic results are obtained under the following simplifying idealizations:
\begin{enumerate}
\item
The derivation is done in the thermodynamic limit, while typical laboratory BECs contain a {\it finite} number of atoms. 
\item
The spacing between energy levels of the Bose gas is assumed to be much smaller than $k_BT$ for the 
temperatures under consideration, so that a sum over 
occupied energy levels can be replaced by an integral over energy. This semi-classical approximation 
is true only for an infinite, uniform system. However, if the condensate is confined in a finite trap 
({\it e.g.} a harmonic trap), the level-spacing becomes appreciable and the approximation loses its 
validity. In the semi-classical approach, population of the lowest state is taken separately outside 
the integral, since the semi-classical density of states vanishes, while the Bose distribution function 
diverges at the lower limit. 
\item
The chemical potential ($\mu$) is assumed to remain zero for $T \leq T_c^0$. However, when the sum is 
evaluated exactly for a finite $N$, $\mu$ (determined from the condition that the total number of 
bosons is $N$) decreases with increase of $T$, initially slowly from zero at $T=0$ up to $T=T_c$ and 
then rapidly for $T>T_c$. 
\item
Interatomic interactions are disregarded. In reality atoms interact through well known interatomic 
interactions. At the very low energy and temperature of a BEC, the $s$-wave scattering length ($a_s$) 
governs the effective interaction. The latter can be repulsive as in $^{87}$Rb and $^{23}$Na atoms, or 
attractive as in $^{7}$Li atoms. It is also possible to `tune' the interaction to any desired value 
using the Feshbach resonance~\cite{Roberts}. The interactions have a profound effect on the condensate 
properties, as also on $T_c$~\cite{Dalfovo}. In a harmonically trapped repulsive BEC, the hydrodynamic 
model shows that energy of most levels are lower than those of the corresponding non-interacting 
bosons~\cite{Stringari}. Thus smaller thermal energy is  needed to lift particles from the ground 
state. Consequently the critical temperature reduces appreciably. 
\end{enumerate}

Attempts have been made to remove some of these simplifying idealizations. Replacing the 
sum by an integration including semi-classical density of states, Bagnato and Kleppner~\cite{Bagnato} 
showed that BEC is possible in one-dimension, if the trap is more confining than a parabolic potential, 
while in two-dimension, BEC is possible for {\it any power law} confining potential. Using a modified 
semi-classical approximation, Yukalov~\cite{Yukalov} replaced the lower limit of integration from zero 
to a finite value given by the uncertainty relation. He thus removed the divergence at the lower limit 
of the integration to obtain Bose condensation for an arbitrary power law confinement in any dimension. 
Ketterle and van Druten~\cite{Ketterle} obtained analytic expressions for the condensate fraction and 
$T_c$, after relaxing the first two conditions: treating the sums appropriately for a finite $N$ and 
using some mathematical approximations. They demonstrated that BEC is possible in three, two and even one 
dimensional harmonic traps. This is in sharp contrast with the semi-classical treatment. The critical 
temperature in one dimension is higher than in three dimensional traps and therefore more convenient 
for experimental achievement of BEC in quasi-one dimensional traps. However, they still assumed 
$\mu=0$ for $T \leq T_c$ and considered only non-interacting bosons.  All these attempts still retain 
some of the simplifying assumptions. Moreover, it is not always possible to get analytic expressions 
valid at $T_c$, approaching it both from below as also from above. \\

In the exact treatment of finite systems, in which sums are evaluated exactly, all thermodynamic quantities 
become continuous functions of $T$~\cite{Napolitano, Biswas}.  Consequently, {\it there is no strictly critical 
temperature} in such a system, although for large enough $N$, there is a distinct change in the nature of the 
curves over a small interval around a particular temperature, referred to as the transition temperature ($T_c$). 
For very large $N$, the first derivatives tend to exhibit a discontinuity at $T_c$. Thus in the strict sense, 
there is no phase transition in a finite inhomogeneous system. 
The effects of semi-classical approximations in infinite systems and disregard 
of inter-particle interactions make $T_c^0$ appreciably larger than $T_c$. \\

For a true phase transition, one can define {\it critical exponents}, which characterize the nature of phase 
transition and its 
universal properties~\cite{Yeomans}. When the transition from the condensed phase to the Bose gas phase is 
gradual and there 
is no true phase transition, the critical exponents vanish. Still, for a large enough $N$. there is a sharp 
enough transition, which we refer to as {\it quasi phase transition} (QPT). In this case, we define a {\it 
transition exponent} (TE), which shows the characteristic nature of the thermodynamic quantity across the transition 
temperature. In this work, we evaluate and examine the transition exponents (defined in the next Section) to provide an 
understanding of the nature of thermodynamical quantities near a quasi phase transition. Since a laboratory 
condensate contains interacting 
atoms, we also investigate interacting condensates. Inclusion of interatomic interactions, makes the many-body 
problem non-trivial. We use the correlated 
potential harmonic expansion method (CPHEM)~\cite{Das2} for solving the interacting many-body system. \\

The paper is organized as follows. Sec. II presents the theoretical background, providing definitions and basic equations. 
Sub-section II.A defines the transition exponent for a quasi phase transition. In the next sub-section we discuss how 
thermodynamic quantities and TE can be evaluated for a finite system. A third sub-section is included to 
briefly outline the CPHEM. 
Sec. III presents the results of our calculation and finally we draw our conclusions after a brief summary in Sec. IV.

\section{Theoretical background}

\subsection{Transition exponent for a quasi phase transition}

\hspace*{1cm}
An important physical quantity in connection with a true phase transition at a critical point is the critical exponent. 
A critical point is characterized by divergences or discontinuities in thermodynamic quantities, depending on 
the nature of the thermodynamic quantity and dimensionality of the system. 
Thus the thermodynamic quantity may cease to be analytic at the critical point. The critical exponent is useful 
in understanding 
the rapidly changing behavior of thermodynamic functions at the critical temperature ($T_c^0$). In terms of a 
reduced temperature
\begin{equation}
t=\frac{T-T_c^0}{T_c^0},
\label{Tcspht}
\end{equation}
which is a dimensionless measure of the deviation of temperature from the critical 
temperature, one can define the critical exponent ($\lambda$) for a
thermodynamic function $F(t)$ as~\cite{Yeomans} 
\begin{equation}
\lambda=\lim_{t\rightarrow 0} \frac{{\rm ln}|F(t)|}{{\rm ln}|t|}.
\label{lambda}
\end{equation}
This corresponds to $F(t)\sim |t|^{\lambda}$, for small values of $|t|$.
Obviously, at the critical
temperature, $F(t)$ must either vanish or be singular, depending on the sign
of $\lambda$ (for $\lambda \neq 0$). But sometimes (particularly in a QPT) thermodynamic quantities are neither singular, 
nor do they vanish at the transition temperature. A typical example 
is the heat capacity of a BEC, either a finite one or in the thermodynamic limit.  In such cases, one can take 
$F(t)$ to be the
difference of the desired thermodynamic function at $T$ and its value at $T_c$. 
Indeed, in this case, $\lambda$ is zero and the following prescription will give 
$\lambda_1$ (which must be positive) of the relation~\cite{Yeomans}
\begin{equation}
F(t)=F(0)+b|t|^{\lambda_1}+... \hspace*{0.2cm}.
\end{equation}
Since in this case, $\lambda$ is zero, $\lambda_1$ is the leading exponent 
of interest. The nature of the thermodynamic function near the critical 
point is determined by $\lambda_1$. Hence for such a situation, we 
{\it define} $\lambda_1$ as {\it the transition exponent}, given by 
\begin{equation}
\lambda_1=\lim_{t\rightarrow 0} \frac{\ln|F(t)-F(0)|}{\ln|t|} .
\label{TE}
\end{equation}
For $T<T_c^0$, the condensate fraction for non-interacting bosons in the thermodynamic limit 
is $\frac{N_0}{N}=1-(\frac{T}{T_c^0})^{\alpha}$, where 
$\alpha=\frac{3}{2}$ for a uniform system, while $\alpha=3$ for bosons in a 3-D harmonic trap. Then it is easy 
to see that $\lambda=1$. 
For the heat capacity, $C_N(T)/(\alpha N k_B)=a(T/T_c^0)^{\alpha}$ ($a$ being a constant) and one has $\lambda=0$. 
On the other hand, chemical potential remains zero for $T \leq T^0_c$ and for $T>T^0_c$ it is obtained numerically 
even in the thermodynamic limit~\cite{Huang}. Hence no closed analytic form is possible and numerical analysis is 
the only possibility. Thus for chemical potential $\lambda$ is undefined for $T < T^0_c$ and has to be obtained 
numerically for $T>T^0_c$, even in the thermodynamic limit. \\

For QPT with $N$ small, 
the thermodynamic functions are smooth across the transition region. If the function does not have an extremum at 
$T_c$ [like chemical potential and condensate fraction in 1-D, 2-D and 3-D and specific heat in 1-D (see later)] a 
simple Taylor 
series expansion shows that $\lambda=0$ and $\lambda_1=1$. On the other hand heat capacity in 2-D and 3-D 
(for both interacting and 
non-interacting bosons) has a maximum and one sees that $\lambda=0$ while $\lambda_1=2$. But in the 
$N \rightarrow \infty$ limit, 
exact numerical calculation shows a {\it sharp fall} in the value of $C_N(T)$. For a mathematical discontinuity, 
$\lambda_1$ 
may be different from $2$, and may have different values for $T<T_c$ and for $T>T_c$. Thus it is interesting to 
calculate $\lambda_1$ numerically for a large 
enough value of $N$, to understand how the thermodynamic function changes across the transition region, as $N$ 
increases. \\

An important characteristic property is the universality of the critical
exponent. The value of $\lambda$ does not depend on the interatomic
interaction  or detailed nature of the system. Its value depends only on
the dimension of the system and the symmetry of the order 
parameter~\cite{Yeomans}. We expect a similar universality property satisfied by the transition exponents 
($\lambda_{1}$) as well. In the
present work, we calculate transition exponents for a number of thermodynamic
functions of the BEC in different cases and investigate whether they depend
on the system parameters. We consider a non-interacting Bose gas trapped in a 
harmonic oscillator potential in one, two and three dimensions. Finally, we 
also include realistic interatomic interactions in an approximate many-body 
treatment of the real Bose gas. As discussed 
earlier, thermodynamic functions like specific heat, condensate fraction 
and chemical potential depend on the dimension of the system, 
the choice of the trap potential, whether interatomic interactions are 
included or not, {\it etc}. But transition exponents extracted from them may display a universal property, 
similar to that exhibited by the critical exponents in a true phase transition. 
Our present work aims to explore whether the transition exponents exhibit an underlying universality. \\

\subsection{System of non-interacting bosons}

\hspace*{1cm}
In this sub-section, we consider $N$ non-interacting bosonic atoms trapped in an isotropic harmonic 
potential of frequency $\omega$ in $d$-dimensional space ($d=1,2,3$). The energy scale is so chosen that the single 
particle ground state is at zero energy. The energy eigenvalues 
$E_{n}$ $(n=0,1,2,...)$ are given by
\begin{equation}
 E_{n} = n\hbar\omega
\label{energy}
\end{equation}
The number of particles in the $n$-th state with energy $E_n$ at 
a temperature $T$ is given by the Bose distribution function 
\begin{equation}
 f(E_{n}) = \frac{1}{e^{\beta(E_{n}-\mu)} -1}
\label{distribution}
\end{equation}
where $\beta = 1/{k_{B}}T$ and $\mu$ is the chemical potential. The latter 
is determined from the constraint that the total number of particles is $N$
\begin{equation}
 N = \sum_{n=0}^{+\infty} \gamma_{n}f(E_{n}), 
 \label{sum}
\end{equation}
where $\gamma_n$ is the degeneracy of the $n$-th level. It is $1$, $(n+1)$ and $\frac{(n+1)(n+2)}{2}$ for the one-, 
two- and three-dimensional harmonic oscillator respectively. Clearly, $\mu$ has a temperature dependence. 
The total energy for the system is given by 
\begin{equation}
 E(N,T)=\sum_{n=0}^{+\infty} \gamma_{n}f(E_{n})E_{n}
 \label{sumen}
\end{equation}
The specific heat for fixed particle number ($N$) is calculated using the relation
\begin{equation}
 C_{N}(T) = \frac{\partial E(N,T)}{\partial T}
 \label{spheat}
\end{equation}
Using (\ref{distribution}), (\ref{sumen}), (\ref{spheat}), one can obtain the 
heat capacity as 
\begin{eqnarray}
 & C_{N}(T) = \beta\sum_{n=0}^{+\infty} \frac{\gamma_{n}E_{n}\exp{(\beta(E_{n}-\mu))}}{(\exp{(\beta(E_{n}-\mu))}-1)^2}[\frac{E_{n}-\mu}{T}+\frac{\partial \mu}{\partial T}] \nonumber \\
 & =\beta\sum_{n=0}^{+\infty}  \frac{\gamma_{n}E_{n}\exp{(\beta(E_{n}-\mu))}}{(\exp{(\beta(E_{n}-\mu))}-1)^2}[\frac{E_{n}-\mu}{T}- \frac{\sum_{m=0}^{+\infty}\gamma_{m}(E_{m}-\mu)\exp{(\beta(E_{m}-\mu))}
(f(E_{m}))^2}{T\sum_{p=0}^{+\infty}\gamma_{p}\exp{(\beta(E_{p}-\mu))}(f(E_{p}))^2}].
\label{spheatform}
\end{eqnarray} 
\hspace*{1cm}

In the semi-classical treatment, $\mu$ is assumed to remain zero for $T \leq T^0_c$ and start decreasing for 
$T>T^0_c$. But a numerical solution of Eq.~(\ref{sum}) for a {\it finite} $N$  shows that $\mu$ decreases 
{\it very slowly} from its maximum value (zero) at $T=0$, as $T$ increases in the interval $T \leq T_c$. 
The rate of decrease becoming {\it suddenly rapid} at the transition temperature $T_c$. Thus, in this case 
there is no sharp critical temperature. 
The heat capacity also becomes a smooth function of $T$, attaining a maximum at 
a temperature, at which $\mu$ suddenly becomes a rapidly decreasing function of $T$ (except for the 
one-dimensional case, see below). 
The {\it transition temperature} $T_c$ is usually defined as the temperature at which 
$C_N(T)$ is a maximum~\cite{Napolitano}
\begin{equation}
\frac{\partial C_N(T)}{\partial T}{\Big |}_{T_c}=0.
\label{defTc}
\end{equation}

We use our definition of TE, Eq.~(\ref{TE}), separately for $T>T_c$ and $T<T_c$ (denoted by the 
superscripts $+$ and $-$ respectively), for different thermodynamic functions ($X$) 
\begin{equation}
\lambda_1^{X,\pm}=\lim_{t \rightarrow 0\pm}\frac{\ln |F^X(t)-F^X(0)|}{\ln |t|},
\label{TE-gen}
\end{equation}
where $t=(T-T_c)/T_c$ and $X$ stands for chemical potential (chempot), condensate fraction (condfrac) 
and specific heat (spht) and $F^X(t)$ is $\mu(t)$, $\frac{N_0}{N}(t)$ and $C_N(t)$ respectively. For 
all thermodynamic functions, we use the 
same transition temperature defined by Eq.~(\ref{defTc}). 

\subsection{Realistic interacting bosons}

For the interacting case, one has to solve the many-body Schr\"odinger equation. 
An essentially exact solution is possible by the diffusion Monte Carlo (DMC) method~\cite{Blume}. However, 
this has been done only for $N$ less than $\sim 100$. The mean-field approach together 
with the assumption of a contact two-body interaction leads to the commonly 
used Gross-Pitaevskii equation (GPE)~\cite{Dalfovo}. In this approach, 
all correlations 
are disregarded and no realistic two-body interaction can be used. In addition, 
for an attractive BEC, there is a pathological singularity at the 
origin~\cite{Kundu}. Hence, we adopt a simplified few-body technique, called 
correlated potential harmonics expansion method (CPHEM)~\cite{Das1,Das2}, in which all two-body 
correlations are retained, but higher-body correlations are neglected. Disregard of higher-order correlations 
is manifestly justified for a laboratory BEC, since it is designed to be so dilute that three-body collisions 
do not take place, to preclude formation 
of molecules and consequent depletion of the condensate through thee-body recombination. Furthermore, {\it any realistic} 
interatomic interaction can be incorporated in the CPHEM. This technique has been 
successfully applied to both repulsive and attractive BECs~\cite{PHEM}. \\
\hspace*{1cm}
In the following, we review the CPHEM very briefly. Interested readers can 
get the details from Refs.~\cite{Das1,Das2}. Schr\"odinder equation for the relative 
motion of a system of $N$ identical spinless bosons, 
interacting via pair-wise potential $V$ and trapped by $V_{trap}$ is 
\begin{eqnarray} 
\Big[-\frac{\hbar^{2}}{m} \sum_{i=1}^{\cal {N}} \nabla_{\vec{\zeta}_{i}}^{2}+
V_{trap}(\vec{\zeta}_1, ..., \vec{\zeta}_{\cal {N}}) + 
V(\vec{\zeta}_{1}, ..., \vec{\zeta}_{\cal{N}})-E_{R} \Big] 
\psi(\vec{\zeta}_{1}, ..., \vec{\zeta}_{\cal{N}}) = 0 ,
\label{SE_rel_motion}
\end{eqnarray} 
where $\{ \vec{\zeta}_1, \vec{\zeta}_2, \dots, \vec{\zeta}_{\cal{N}} \}$ is the set of ${\cal{N}}=(N-1)$  
Jacobi vectors, which are the relative variables, after separation of the center of mass 
motion~\cite{Ballot} and $E_R$ is the energy of the relative motion. A global length called `hyperradius' is defined as 
\begin{equation}
r=\left[\sum_{i=1}^{\cal N} \zeta_i^2\right]^{\frac{1}{2}}.
\end{equation}
This, together with a set of $(3{\cal N}-1)$ `hyperangles', constituted by $2{\cal N}$ polar angles of ${\cal N}$ Jacobi 
vectors and $({\cal N}-1)$ angles defining their relative 
lengths~\cite{Ballot} define the hyperspherical variables, replacing the Jacobi vectors. These are 
$3{\cal{N}}$-dimensional analogue of 
$3$-dimensional spherical polar coordinates. Likewise, the $3{\cal{N}}$-dimensional analogue of spherical 
harmonics are the 
hyperspherical harmonics (HH). These are the eigenfunctions of the grand orbital operator, which is the 
hyperangular part of 
the $3\cal{N}$-dimensional Laplace operator $\sum_{i=1}^{\cal {N}} \nabla_{\vec{\zeta}_{i}}^{2}$~\cite{Ballot}. 
It is natural 
to expand $\psi$ in the complete set of HH, giving rise to the 
hyperspherical harmonics expansion method (HHEM). But the degeneracy of the HH basis increases very rapidly 
with $N$.  Consequently, 
imposition of symmetry and calculation of matrix elements become practically impossible for $N>3$. Use of the 
full HH basis for the expansion 
of $\psi$ includes all many-body correlations in the wave function. However as mentioned earlier, a typical 
laboratory BEC is designed 
to be physically very dilute, since otherwise three-body collisions will lead to formation of molecules and 
consequent depletion 
of the condensate~\cite{Dalfovo}. This means that three- and higher-body correlations are negligible in such 
condensates. 
Hence $\psi$ can be decomposed into interacting-pair Faddeev components, $\psi_{ij}$ (which becomes a 
function of the $(ij)$-pair separation $\vec{r}_{ij}$ and hyperradius $r$ only, due to neglect of higher than 
two-body correlations) 
\begin{equation}
\psi=\sum_{i,j>i}^N\psi_{ij}(\vec{r}_{ij},r).
\end{equation}
Then, instead of the full HH basis, one can choose a subset, called potential harmonics (PH) subset~\cite{Fabre} 
for the expansion of $\psi_{ij}$. The PH subset is defined 
as the subset of HH necessary for the expansion of the two-body interaction, $V(\vec{r}_{ij})$. Since 
$\psi_{ij}$ is a function of $\vec{r}_{ij}$ and $r$ only, the PH basis is sufficient for its expansion, which reads 
\begin{equation}
 \psi_{ij}(\vec{r}_{ij},r)=r^{-\frac{(3{\cal N}-1)}{2}}
\sum_{K}\mathcal{P}_{2K+l}^{lm}(\Omega^{ij}_{\cal N})u_{K}^{l}(r), 
\label{PHexpn}
\end{equation}
where $\mathcal{P}_{2K+l}^{lm}(\Omega^{ij}_{\cal N})$ is a potential harmonic~\cite{Fabre}. 
The $r$-dependent factor in front is included to remove the first  derivative with respect to $r$. 
Although each HH is in general a function of all $3{\cal {N}} - 1$ hyperangles, $\Omega^{ij}_{\cal N}$, the 
PH, being a subset of HH sufficient for the expansion of $V(\vec{r}_{ij})$, is a function of only three 
hyperangles: polar angles of $\vec{r}_{ij}$ and a hyperangle ($\phi$) defined through $r_{ij}=r\cos \phi$. 
Corresponding quantum numbers are $l$, $m$ and $K$. This corresponds physically to freezing all 
irrelevant degrees of freedom, and setting corresponding quantum numbers to zero. The physical 
picture is that when the $(ij)$-pair interacts, rest of particles are inert spectators and do 
not contribute to orbital and grand-orbital angular momenta. Thus the orbital angular momentum of 
the system is contributed by the interacting pair only~\cite{Das1,Das2}.
Substitution of the expansion, Eq.~(\ref{PHexpn}) in 
the Faddeev equation for  the $(ij)$-partition 
\begin{equation}
 {\Big (}-\frac{\hbar^{2}}{m}\sum_{i=1}^{\cal N} \nabla^{2}_{\vec{\zeta}_{i}}+V_{trap}-E_{R}{\Big )}\psi_{ij}=-V(r_{ij})\sum_{k,l>k}^{N} \psi_{kl},
\label{Faddeveq}
\end{equation}
and projection on the PH corresponding to the $(ij)$-partition 
give a set of coupled differential equations in $r$. Any suitable interatomic potential 
can be chosen for $V(\vec{r}_{ij})$. Realistic potentials have a strong 
repulsion at very short separations. Hence $\psi_{ij}$ should be extremely  
small at such separations. On the other hand, the leading potential harmonics 
(corresponding to $K=0$ and small $K$ values) of Eq.~(\ref{PHexpn}) are appreciably 
large for small $r_{ij}$, resulting in a very slow rate of convergence. This is 
corrected by inclusion of an additional correlation function, $\eta(\vec{r}_{ij})$, which simulates the 
nature of $\psi_{ij}(\vec{r}_{ij},r)$ for small $r_{ij}$
\begin{equation}
 \psi_{ij}(\vec{r}_{ij},r)=r^{-\frac{(3{\cal N}-1)}{2}}
\sum_{K}\mathcal{P}_{2K+l}^{lm}(\Omega^{ij}_{\cal N})u_{K}^{l}(r)
\eta(\vec{r}_{ij}).
\label{CPHexpn}
\end{equation}
Note that the energy of the interacting pair in the condensate is negligibly small 
compared with the energy scale of the interatomic interaction. Hence at small values 
of $r_{ij}$, $\psi_{ij}$ should behave as the zero energy solution, $\eta(\vec{r}_{ij})$,  of the interacting pair
\begin{equation}
-\frac{\hbar^2}{m}\frac{1}{r_{ij}^2}\frac{d}{dr_{ij}}\left(r_{ij}^2
\frac{d\eta(r_{ij})}{dr_{ij}}\right)+V(r_{ij})\eta(r_{ij})=0. 
\label{2Beqn} 
\end{equation}
With the addition of this intuitive {\it short-range correlation function}, rate of 
convergence is dramatically improved. Moreover, the asymptotic form of 
$\eta(\vec{r}_{ij})$ is $C(1-a_s/r_{ij})$, which depends on the $s$-wave scattering 
length ($a_s$)~\cite{Pethick}. Hence the short-range repulsion of the realistic two-body 
potential can be adjusted to correspond to the appropriate $a_s$ (specifying the effective 
two-body interaction) for the chosen condensate. \\

Substitution of the expansion, Eq.~(\ref{CPHexpn}) in 
Eq.~(\ref{Faddeveq}) followed by projection on the PH corresponding to the 
$(ij)$-partition give a set of coupled differential equations (CDE) 
\begin{eqnarray}
\Big[-\frac{\hbar^{2}}{m} \frac{d^{2}}{dr^{2}} + \frac{\hbar^{2}}{mr^{2}}
\{ {\cal L}({\cal L}+1) &+& 4K(K+\alpha+\beta+1)\} 
+ V_{trap}(r) - E_R  \Big] U_{Kl}(r) \nonumber \\
&+& \sum_{K^{\prime}}f_{Kl}V_{KK^{\prime}}(r)f_{K'l} U_{K^{\prime}l}(r) = 0,
\label{CDE}
\end{eqnarray}
where $U_{Kl}(r) = f_{Kl}u_{K}^{l}(r)$,  ${\cal L} =
l+\frac{3N-6}{2}$, $\alpha=\frac{3N-8}{2}$, $\beta=l+\frac{1}{2}$.  
$l$ is the orbital angular momentum of the condensate. The constant $f_{Kl}^2$ is  
the overlap of the PH for interacting partition with the sum of PHs corresponding to 
all partitions~\cite{Fabre}. The correlated potential matrix 
element $V_{KK^{\prime}}(r)$ is given by~\cite{Das2} 
\begin{eqnarray}
\hspace*{-0.5cm}V_{KK^{\prime}}(r) = (h_{K}^{\alpha\beta}
h_{K^{\prime}}^{\alpha\beta})^{-\frac{1}{2}}
\int_{-1}^{+1} 
P_{K}^{\alpha 
\beta}(z)
V\left(r\sqrt{\frac{1+z}{2}}\right) 
P_{K^{\prime}}^{\alpha \beta}(z)\eta\left(r\sqrt{\frac{1+z}{2}}\right)
W_{l}(z) 
dz,
\label{corrPME}
\end{eqnarray}
where $h_{K}^{\alpha\beta}$ and $W_{l}(z)$ are respectively the norm
and weight function~\cite{Abramowitz} of the Jacobi polynomial
$P_{K}^{\alpha \beta}(z)$. 
Since $\eta(\vec{r}_{ij})$ is included, the expansion basis is no longer orthogonal. 
One can follow standard procedure for non-orthogonal basis. However, dependence on 
$r$ of the overlap matrix complicates this procedure. On the other hand, we found 
that $\eta(r_{ij})$ obtained numerically from Eq. (\ref{2Beqn}) differs appreciably 
from a constant value only in a very small interval of small $r_{ij}$ values. Hence 
$<{\mathcal P}^{lm}_{2K+l}(\Omega^{(ij)}_N)|
{\mathcal P}^{lm}_{2K+l}(\Omega^{(kl)}_N)\eta(r_{kl})>$ is nearly independent of 
$r$. Disregarding its derivatives we approximately get Eq. (\ref{CDE}), 
with $V_{KK^{\prime}}(r)$ given by Eq. (\ref{corrPME}). The fact that  
the overlap is not one is taken through the asymptotic constant $C$ of 
$\eta(r_{ij})$. This implies that pairs of atoms at very low energy interact via an 
effective interaction $V(r_{ij})\eta(r_{ij})$. This can be understood as follows. 
Atoms having a very large de Broglie wave length at extremely low energy cannot come 
too close to feel the actual atom-atom potential, which is very strong. In this limit 
the total scattering cross section is $4\pi|a_s|^2$ and $a_s$ specifies the {\it effective 
atom-atom interaction} in the zero energy limit~\cite{Pethick,Dalfovo}. A fairly fast computer code can solve 
Eq. (\ref{CDE}) using the hyperspherical adiabatic approximation~\cite{Das3} with upto 
15000 particles in the condensate. This technique has been tested against known results, 
both experimental ones and theoretical 
ones calculated by other authors, for repulsive as well as attractive 
condensates~\cite{Das1,Das2,PHEM}. The realistic van der Waals (vdW) potential is used 
to obtain a large number of energy 
eigenvalues of the entire BEC for different orbital angular momenta 
of the system. These are then used in Eq.(\ref{sum}) to calculate the 
chemical potential at a chosen temperature $T$. 
Note that in this case, the sums over $n$ in Eqs,(\ref{sum}), (\ref{sumen}) 
and (11)
are to be replaced by double sums over $\{n,l\}$ and $\gamma_n$ 
is replaced by  
$(2l+1)$. The energy eigenvalue $E_{nl}$ is the energy of the 
$n$-th excitation for the orbital angular momentum $l$ of the 
system. 
Finally Eq.~(11) 
is used to calculate $C_N(T)$~\cite{Biswas}. Condensate fraction is 
obtained as $\frac{N_0}{N}$, where $N_0$ is the ($n=0$, $l=0$) term 
of Eq.(\ref{sum}). \\

\section{Results}

\subsection{Non-interacting bosons}

\hspace*{1cm}
In this sub-section, we consider $N$ non-interacting bosons in an isotropic harmonic trap 
in one, two and three dimensions.
We first present the results for the one dimensional case. 
A plot of condensate fraction against $k_BT/\hbar\omega$ shows a fast and almost linear drop for a considerable 
stretch, followed by a slowly decreasing part gradually going to zero as is shown in Fig.~1. The change over from the sharp linear drop 
to the gradually decreasing portion is fairly sudden for large $N$. This shows that a QPT occurs, in agreement 
with Ketterle and van Druten~\cite{Ketterle}, and in contradiction with semi-classical treatment in text books. 
In Ref.~\cite{Ketterle}, chemical potential was taken to be zero for $T \leq T_c$. This is strictly true in the 
thermodynamic limit only. When $N$ is finite, $\mu$ {\it decreases slowly} for $T \leq T_c$, then rapidly for 
$T>T_c$. In this case, $C_N(T)$ is a monotonically increasing function (see below). 

\begin{figure} [ht]
\noindent \includegraphics[clip,width= 16cm,height=9cm,
angle=0]{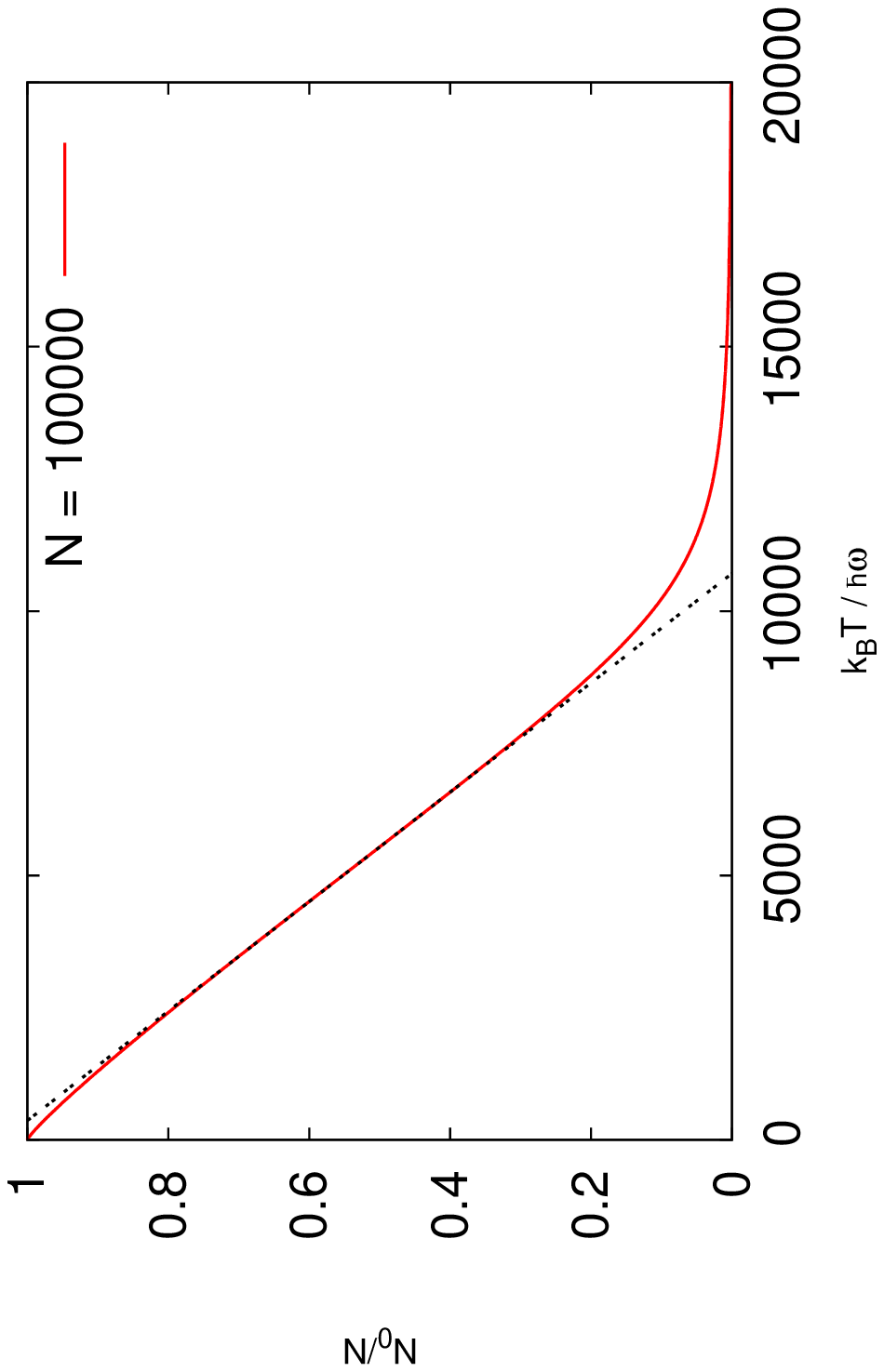}
\caption{(Color online) Plot of condensate fraction as functions of $k_{B}T/\hbar\omega$ for 
$100000$ non-interacting bosons in a one-dimensional
harmonic trap. The extrapolated straight portion 
with the horizontal axis of condensate fraction versus $k_BT/\hbar\omega$ plot,as shown by the black 
dotted line gives the transition temperature $k_BT_c/\hbar\omega$.}
\end{figure}

In the absence of definition 
Eq.~(\ref{defTc}), 
we take the transition temperature ($k_BT_c/\hbar\omega$) to be the intercept of the extrapolated straight portion 
with the horizontal axis of condensate fraction versus $k_BT/\hbar\omega$ plot. 
This value is somewhat larger than that in Ref.~\cite{Ketterle}. In Table~I, we present 
calculated values of $k_BT_c/\hbar\omega$ for different $N$, together with the the values obtained from a numerical 
solution of Eq.~(16) of Ref.~\cite{Ketterle}. It is seen that the percentage difference is fairly large for small 
$N$ and decreases with increasing $N$. \\
\begin{table}[h!]
\caption{ BEC transition temperature ($T_c$) for non-interacting bosons in one dimensional trap.}
\begin{center}
\begin{tabular}{|c|c|c|}\hline
$N$ & \multicolumn{2}{c|}{$k_BT_c/\hbar\omega$} \\\cline{2-3}
 & Calculated & From Ref.~\cite{Ketterle} \\\cline{1-3}
1000 & 196.0 & 171.3 \\
10000 & 1381.1 & 1274.9 \\
100000 & 10713.6 & 10088.4 \\\hline
\end{tabular}
\end{center}
\end{table}

We next present the thermodynamic quantities as functions of $T/T_c$. 
\begin{figure} [ht]
\noindent \includegraphics[clip,width= 16cm,height=9cm,
angle=0]{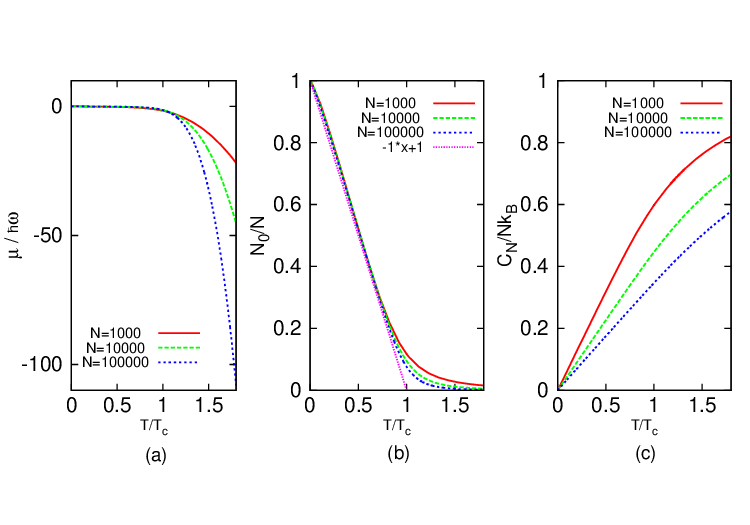}
\caption{(Color online) Plot of chemical potential (panel a), 
condensate fraction (panel b) 
and $C_N/Nk_B$ (panel c) as functions of $T/T_c$ for 
indicated number ($N$) of non-interacting bosons in a one-dimensional
harmonic trap. Note that chemical 
potential is expressed in energy oscillator unit ($\hbar\omega$).
Note also that $C_N/Nk_B$ is a {\it monotonically increasing} 
function of $T$, giving 
rise to the common notion that there is no criticality in one dimension.}
\end{figure}
In panel (a) of Fig.~2, we 
plot $\mu/\hbar\omega$ against $T/T_c$ for $N= 
1000$, $10000$ and $100000$. One notices that as temperature increases, $\mu$ remains nearly constant for 
$T/T_c$ less than about $1$, after which it decreases rapidly. The change over in the nature of the decrease 
around $T_c$ becomes sharper as $N$ increases. This again clearly demonstrates the occurrence of a QPT at $T_c$. 
In panel (b) of Fig.~2, we plot the condensate fraction,  
as a function of $T/T_c$ for the chosen values of $N$. 
As stated earlier, we notice that a rapid change in the rate of decrease occurs at around $T_c$, the change 
being sharper as $N$ increases. This demonstrates that a BEC phase is possible and it goes gradually over to 
the normal Bose gas phase. In panel (b), we also include the straight line $-1*x+1$, with $x=T/T_c$. The 
curves overlap more and more with this straight line over a considerable region as $N$ increases. However, 
from a plot of $C_N/Nk_B$ 
against $T/T_c$ in 
panel (c), we notice that {\it for the one-dimensional case}, $C_N(T)$ is a monotonically increasing function 
of $T$
This implies that the 
one-dimensional case is distinctly different from higher dimensions. However, we notice from Fig.~2(c), 
that $C_N(T)$ is almost linear for $T$ far away from the transition region, with distinctly different slopes. 
The difference of slopes decreases as $N$ increases. Consequently, criticality in the nature of $C_N(T)$ gets 
smeared out as $N$ increases for one dimensional case. \\
\begin{figure}[ht]
\noindent \includegraphics[clip,width= 16cm,height=9cm,
angle=0]{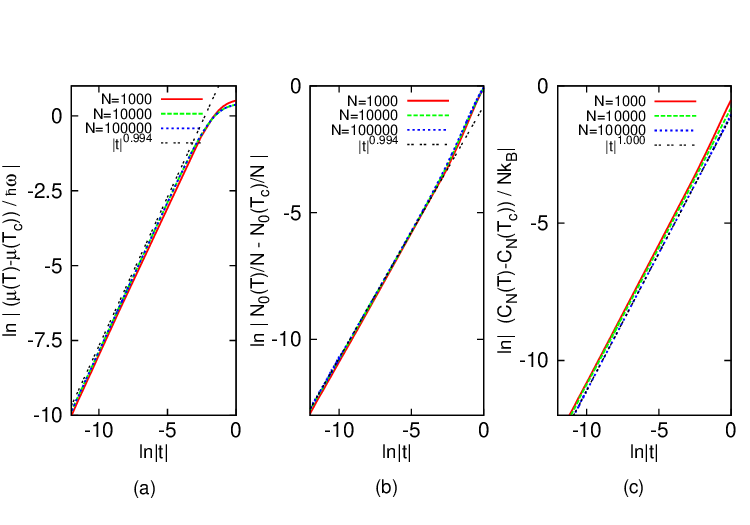}
\caption{(Color online) Plot of $\ln|(\mu(T)-\mu(T_c))/\hbar\omega|$ 
(panel a), 
$\ln|N_0(T)/N-N_0(T_c)/N|$ (panel b)
and $\ln|(C_N(T)-C_N(T_c))/Nk_B|$ (panel c) as functions of $\ln|t|$ for
indicated number ($N$) of non-interacting bosons in a one-dimensional
harmonic trap.
Typical straight line fits to the asymptotic regions are shown 
by black dotted lines, indicated as $|t|^{\lambda_1}$, giving the best fit value of $\lambda_1$. Note 
that the asymptotic linear regions for different $N$ are parallel, 
showing that the transition exponent of a particular thermodynamic
function is independent of $N$.}
\end{figure}

Standard text book treatments, replacing sums by integrals and assuming $\mu=0$ for $T \leq T_c$, in the thermodynamic 
limit, conclude that BEC is not possible in one-dimension. 
Our calculations treating the sums exactly and allowing $\mu$ to take appropriate value at all temperatures, 
show that a QPT is possible even in one 
dimension for a {\it finite} number of bosons. This is in agreement with Ketterle 
and van Druten~\cite{Ketterle} who pointed out that BEC is possible in one dimension at a higher transition 
temperature. But our calculations also show that in the limit $N \rightarrow \infty$, the rapid change in 
$C_N(T)$ across the transition temperature  gradually fades away. \\

Using $T_c$ obtained above, we plot 
$\ln |(\mu(T)-\mu(T_c))/\hbar\omega|$, $\ln |\frac{N_0(T)}{N} - \frac{N_0(T_c)}{N}|$ and $\ln |(C_N(T)-C_N(T_c))/Nk_B|$ 
against $\ln |t|$, for $t<0$, in panels (a), (b) and (c) of Fig.~3 respectively. As expected, 
we obtain straight line fits for small $|t|$ ($\ln|t| < -4$). All such straight line fits 
for different $N$ are parallel (to within numerical errors), 
showing that the transition exponents 
are independent of $N$. A straight line fit for the largest $N$ is 
shown by a black dotted line, identified by $|t|^{\lambda_1}$. The slopes of these lines for panels 
(a), (b) and (c) are respectively $0.994, 0.994$ and $1.000$. 
These are then the $t<0$ transition exponents ($\lambda_1^{X,-}$) for chemical potential, 
condensate fraction and specific heat respectively in the one dimensional harmonic 
oscillator trap. These values and the corresponding TE for $t>0$ are presented in 
Table~II
for the largest $N$ chosen.\\

We next repeat the calculations for a two-dimensional isotropic harmonic 
oscillator trap for $N=1000, 10000, 100000$ and $1000000$. In order to save space, we refrain from presenting 
plots for the 2-D case. In this case, $C_N(T)$ versus $T$ curve has a maximum, 
from which we calculate $T_c$ using Eq.~(\ref{defTc}).  
The asymptotic part of the plots similar to Fig.~3 
are again found to be straight lines, which are parallel for different $N$ of the same panel, indicating that 
the transition exponents are independent of $N$.  
Calculated TE for both $t<0$ and $t>0$ are presented in the second row of Table~II. 

\begin{figure}[ht]
\noindent \includegraphics[clip,width= 16cm,height=9cm,
angle=0]{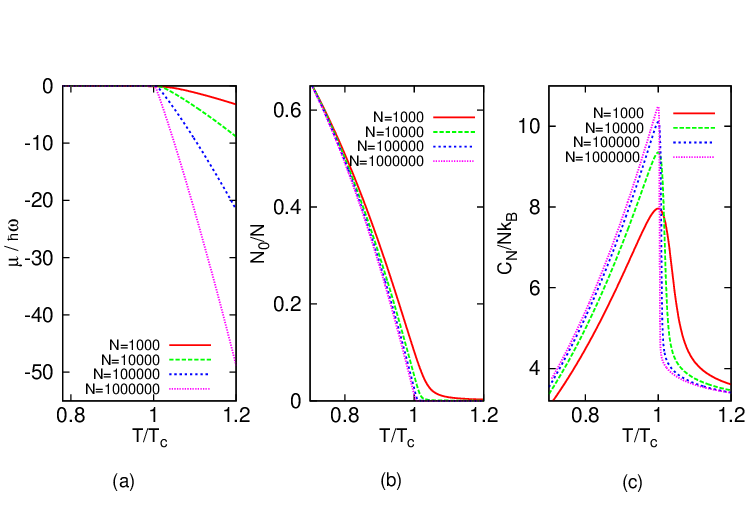}
\caption{(Color online) Plot of $\mu/\hbar\omega$ (panel a), condensate fraction (panel b) and $C_N/Nk_B$ 
(panel c) as functions of $T/T_c$ for
indicated number ($N$) of non-interacting bosons in a
three-dimensional isotropic harmonic trap. In contrast with Fig.~2(c), panel (c) above shows a maximum 
followed by a sharp drop 
in $C_N$, the sharpness of the drop increasing with $N$. 
Furthermore, as $N\rightarrow\infty$, the temperature at which this 
sharp drop occurs approaches $T^0_c$.}
\end{figure}
\hspace*{1cm}

We next calculate these quantities for non-interacting bosons in a three dimensional trap. In this case also, 
$C_N(T)$ shows a maximum, hence we calculate $T_c$ using Eq.~(\ref{defTc}). 
In Fig.~4, we plot $\mu/\hbar\omega$, condensate 
fraction and $C_N/Nk_B$ as functions of $T/T_c$ in panels
(a), (b) and (c) respectively. Calculated values of $T_c/T^0_c$ are $0.896$, $0.949$, $0.976$, and $ 0.989$ 
for $N=1000$, $10000$, $100000$ and $1000000$ 
respectively, where $T^0_c$ is the critical temperature in the semi-classical treatment. The small difference 
between $T_c$ and $T^0_c$ is attributed to the fact that in the exact numerical calculation, $\mu$ is allowed to 
take appropriate values for $T<T_c$, instead of fixing it to be zero for $T \leq T^0_c$. \\

From plots (not presented) similar to those in Fig.~3, we calculate TE for both $t<0$ and $t>0$ for the three 
thermodynamic functions ($X$) for all chosen values of $N$. Once again, we find the linear parts for all $N$ are 
parallel in the case of a given $X$. This again demonstrates that TE for a particular $X$ is independent of $N$. 
Calculated transition exponents for the largest $N$ have been listed in the 
third row of Table~II.\\

In general, it is seen that as $\ln|t|$  
increases, the plots belonging to a particular $X$ for different $N$ separate gradually, 
showing that the nature of the thermodynamic quantity {\it away from the 
QPT region} depends on $N$. 
On the other hand, numerical errors start to show up for large negative $\ln|t|$. 
Hence such points are ignored for calculation of TE. 
Numerical errors are larger 
for the calculation of the specific heat, since it involves many sums, differences and divisions [see 
Eq.~(\ref{spheatform})]. Corresponding plots show some divergences. \\

\begin{table}[h!]
\caption{ Transition exponents for chemical potential, condensate fraction and specific 
heat of BEC for non-interacting bosons in one, two and three dimensions, as also bosons 
interacting through van der Waals potential and trapped by three-dimensional 
harmonic oscillator potential.}
\begin{center}
\begin{tabular}{|c|c|c|c|c|c|c|c|c|c|}\hline
Type of & $N$ & $\frac{k_BT_c}{\hbar\omega}$ & \multicolumn{3}{c|}{$\lambda_1$ for $t<0$} & \multicolumn{3}{c|}{$\lambda_1$ for $t>0$}\\\cline{4-9}
BEC &  &  & $\lambda_1^{\rm chempot,-}$ & $\lambda_1^{\rm condfrac,-}$ & $\lambda_1^{\rm spht,-}$ & $\lambda_1^{\rm chempot,+}$ & $\lambda_1^{\rm condfrac,+}$ & $\lambda_1^{\rm spht,+}$\\\cline{1-9}
1-D &  &  &  &  &  &  &  &  \\
non & $10^5$ & $10713.561$ & $0.994$ & $0.994$ & $1.000$ & $1.007$ & $1.007$ & $0.995$ \\
interacting &  &  &  &  &  &  &  &  \\\cline{1-9}
2-D &  &  &  &  &  &  &  &  \\
non & $10^6$ & $773.259$ & $0.945$ & $1.003$ & $1.790$ & $1.055$ & $0.995$ & $2.153$ \\
interacting &  & & &  &  &  &  &  \\\cline{1-9}
3-D &  &  &  &  &  &  &  &  \\
non & $10^6$ & $93.559$ & $0.926$ & $0.987$ & $1.783$ & $1.079$ & $1.014$ & $2.049$ \\
interacting &  &  &  &  &  &  &  &  \\\cline{1-9}
3-D &  &  &  &  &  &  &  &  \\
interacting & $5\times 10^3$ & $7.650$ & $1.032$ & $1.004$ & $1.878$ &  $1.005$ & $0.991$ & $2.041$  \\
(vdW pot)&  &  &  &  &  &  &  &  \\\cline{1-9}
\end{tabular}
\end{center}
\end{table}

\subsection{Interacting bosons}

\hspace*{1cm}
Next we present the results of our calculation for $N$ bosons interacting 
via the 
\begin{figure}[ht]
\noindent \includegraphics[clip,width= 16cm,height=9cm,
angle=0]{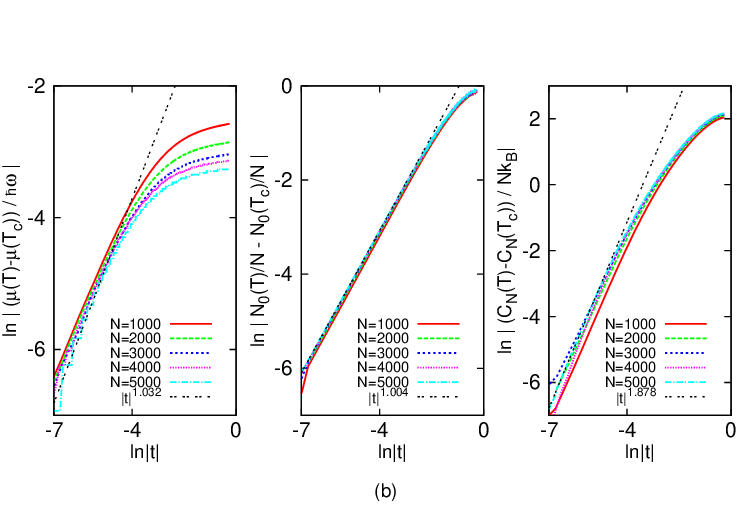}
\caption{(Color online) Plot of $\ln|(\mu(T)-\mu(T_c))/\hbar\omega|$ (panel a),
$\ln|N_0(T)/N-N_0(T_c)/N|$ (panel b) and $\ln|(C_N(T)-C_N(T_c))/Nk_B|$
(panel c) as functions of $\ln|t|$ for
indicated number of atoms interacting via van der Waals potential 
in a three-dimensional harmonic trap.
For other comments, see caption of Fig.~3.}
\end{figure}
van der Waals potential and trapped in an isotropic three 
dimensional harmonic oscillator potential. As mentioned earlier, 
the GPE uses a contact interaction, whose 
strength depends {\it only} on the $s$-wave scattering length 
$a_s$ and hence is shape independent of the two-body potential. 
An earlier calculation~\cite{Chak2} showed that calculated 
observables are indeed {\it shape dependent}. Hence it is appropriate that 
a finite-range realistic interatomic potential like the vdW 
potential should be used.  Calculated values of 
$\ln |(\mu(T)-\mu(T_c))/\hbar\omega|$, $\ln |\frac{N_0(T)}{N} - \frac{N_0(T_c)}{N}|$ and $\ln |C_N(T)-C_N(T_c)|$~\cite{Biswas} 
have been plotted against $\ln |t|$ 
in panels (a), (b) and (c) of Fig.~5. Once again these plots are 
straight lines in the asymptotic ($\ln |t|$ large negative) region. 
The lower limit of $\ln|t|$ has been 
restricted to $-7$ to eliminate numerical errors, which enter inevitably for the interacting case. 
Plots of different $N$, belonging to a particular $X$ are again found to be parallel, 
showing that TE is independent of $N$. Calculated TE are presented in the fourth row of Table~II. \\

Calculation of transition exponent involves logarithms of differences of quantities, for very small 
changes in $T$. Hence, allowing for relatively large errors in the calculation, we see from Table~II that 
the transition exponents for chemical potential and condensate fraction, for both above and below the transition 
temperature are $1$  for all cases studied. It is also $1$ for the heat capacity in the one dimensional 
non-interacting case. For heat capacity in two and three dimensions, the transition exponent is $2$ above 
the transition point and is about $1.8$ below it. We already noticed that TE does not depend on the number of 
bosons in the trap. Furthermore, TEs for interacting bosons in 3D trap are found to be the same 
(within numerical errors) as the corresponding TE for the non-interacting bosons. These show that the 
transition exponents may 
depend on the dimensionality of the system, but not on whether the bosons interact or not. 
Note that although the actual thermodynamic quantities near the transition region 
depend strongly on $N$ and whether the bosons are interacting or not, their intrinsic functional nature 
given by the transition exponent, 
as the transition temperature is approached, has universal characteristics.\\

From actual plot of thermodynamic quantities (see Figs.~2 and 4), one can notice that chemical potential 
and condensate fraction in all cases studied and $C_N(T)$ for one dimensional condensate are smooth 
functions even for very large $N$. A simple calculation shows that TE in such cases should be 1, as we found. 
On the other hand a value of 2 for the TE for $C_N(T)$  (for $T>T_c$ ) in two and three dimensions indicates a 
maximum. For $T<T_c$, TE has a fractional value between 1 and 2. This means that the second derivative of 
$C_N(T)$ appears to diverge at $T \rightarrow T_c$ from below. However the result depends on 
the accuracy of numerical calculation. These behaviors are distinctly different from 
those obtained from the semi-classical treatment.

\section{Conclusions}

\hspace*{1cm}
In conclusion, we remark that a continuous quasi phase transition occurs in 
a non-interacting Bose gas trapped in a harmonic oscillator 
potential {\it even in one-dimension}. This is in sharp contrast with 
standard text book results~\cite{Huang,Pethick}, but in agreement with 
Ketterle and van Druten~\cite{Ketterle}. This is because the replacement 
of sums over single particle states by an integral over energy is not 
a valid approximation for {\it discrete} energy levels, especially when 
$N$ is small. Moreover, the assumption that $\mu=0$ for $T \leq T_c$ is not strictly valid in this case. 
However, we find that there are some distinct 
characteristic features in the one-dimensional case. For example, the 
specific heat is a monotonically increasing function, whereas in higher dimensions it has a maximum. However 
$C_N(T)$ is separately linear with different slopes for $T$ below and above $T_c$, exhibiting a transition 
behavior. But, as $N \rightarrow \infty$, the difference of these slopes tend to vanish. Thus the transition 
characteristics exhibited by the heat capacity gets wiped out in the large $N$ limit. On the other hand, 
chemical potential and condensate fraction continue to show QPT in this limit. 
The dependence of chemical potential, condensate fraction 
and specific heat on temperature for the non-interacting one-dimensionally 
trapped Bose gas has similar behavior as when a Bose gas interacts 
through the harmonic Calogero interaction~\cite{Chak1}.  
The observation that the heat capacity becomes a smooth monotonically increasing function of $T$, with 
difference of slopes gradually decreasing in the large $N$ limit agrees with the commonly accepted idea that there is no 
Bose-Einstein condensation in one-dimension. Ketterle and van Druten 
argued that a quasi one-dimensional experimental setup is advantageous, 
since $T_c$ becomes larger~\cite{Ketterle}. However, in view of the 
above discussion, it is clear that a quasi phase transition in $C_N(T)$ will be obscure in such an experimental 
setup with a large $N$. \\

We have also calculated the transition exponents for three thermodynamic 
functions, {\it viz.} chemical potential, condensate fraction and specific 
heat at constant particle number for a non-interacting Bose gas in an 
isotropic harmonic trap in one, two and three dimensions. For the three 
dimensional case, we have also investigated bosons interacting through a realistic two-body interaction. We find 
that the transition exponent for a particular thermodynamic function does not 
depend on $N$ or whether the bosons are interacting or not. It depends 
on the type of the thermodynamic function 
and the dimensionality of the space. This is consistent with the idea 
of universality. Even though the value of thermodynamic quantities depend strongly on $N$ and whether the 
bosons are interacting or not, besides the dimension of the space, their behavior near the transition point 
has a universal character. It is independent of $N$ and whether or not the bosons interact.\\[2cm]

\hspace*{1cm}
We would like to thank Dr. Parongama Sen for drawing our attention to 
the critical exponent and universality, as well as for useful discussions. 
SG acknowledges CSIR (India) for a Senior 
Research Fellowship (Sanction No. : 09/028(0762)/2010-EMR-I). TKD acknowledges DST (India) for financial assistance through the USERS program.

\end{document}